\documentclass[preprint,aps,12pt,showpacs,nofootinbib,tightenlines]{revtex4}
\usepackage{amsmath}
\usepackage{amssymb}
\usepackage{epsfig}
\usepackage{graphicx}
\begin{document}
\preprint{ZJOU-PHY-TH-07-02}
\preprint{NJNU-TH-07-11}

\newcommand{\beq}{\begin{eqnarray}}
\newcommand{\eeq}{\end{eqnarray}}
\newcommand{\non}{\nonumber\\ }

\newcommand{\acp}{{\cal A}_{CP}}
\newcommand{\etap}{\eta^{(\prime)} }
\newcommand{\etapr}{\eta^\prime }
\newcommand{\jpsi}{ J/\Psi }

\newcommand{\pb}{\phi_{B_s}}
\newcommand{\pp}{\phi_{\pi}}
\newcommand{\pe}{\phi_{\eta}^A}
\newcommand{\pepr}{\phi_{\eta'}^A}
\newcommand{\ppp}{\phi_{\pi}^P}
\newcommand{\pep}{\phi_{\eta}^P}
\newcommand{\peprp}{\phi_{\eta'}^P}
\newcommand{\ppt}{\phi_{\pi}^t}
\newcommand{\pet}{\phi_{\eta}^T}
\newcommand{\peprt}{\phi_{\eta'}^T}
\newcommand{\fb}{f_{B_s} }
\newcommand{\fpi}{f_{\pi} }
\newcommand{\feta}{f_{\eta} }
\newcommand{\fetap}{f_{\eta'} }
\newcommand{\rpi}{r_{\pi} }
\newcommand{\re}{r_{\eta} }
\newcommand{\rep}{r_{\eta'} }
\newcommand{\mb}{m_{B_s} }
\newcommand{\mop}{m_{0\pi} }
\newcommand{\moe}{m_{0\eta} }
\newcommand{\moep}{m_{0\eta'} }

\newcommand{\psl}{ P \hspace{-1.8truemm}/ }
\newcommand{\nsl}{ n \hspace{-2.2truemm}/ }
\newcommand{\vsl}{ v \hspace{-2.2truemm}/ }
\newcommand{\epsl}{\epsilon \hspace{-1.8truemm}/\,  }

\def \epjc{ Eur.Phys.J. C }
\def \jpg{  J. Phys. G }
\def \npb{  Nucl. Phys. B }
\def \plb{  Phys. Lett. B }
\def \pr{  Phys. Rep. }
\def \prd{  Phys. Rev. D }
\def \prl{  Phys. Rev. Lett.  }
\def \zpc{  Z. Phys. C  }
\def \jhep{ J. High Energy Phys.  }

\title{A Study of $B_{d}^0 \to J/\Psi \eta^{(\prime)}$ Decays in the pQCD Approach}
\author{Xin Liu$^{a}$\footnote{ liuxin@zjou.edu.cn},
Zhen-Jun Xiao$^{b}$\footnote{ xiaozhenjun@njnu.edu.cn}, Hui-Sheng Wang$^{c}$}
\affiliation{ $a.$ Department of Physics, Zhejiang
Ocean University, Zhoushan, Zhejiang 316000, P.R. China }
\affiliation{ $b.$ Department of Physics and Institute of Theoretical
Physics, Nanjing Normal University, Nanjing, Jiangsu 210097, P.R. China}
\affiliation{ $c.$ Department of Applied Mathematics and
Physics, Anhui University of Technology and Science, Wuhu, Anhui 241000, P.R. China }  
\date{\today}
\begin{abstract}
Motivated by the very recent measurement of the branching ratio of ${B_d^0} \to J/\psi
\eta$ decay, we calculate the branching ratios of ${B_d}^0
\to J/\psi \eta$ and ${B_d}^0 \to J/\Psi  \eta'$ decays in the perturbative QCD (pQCD) approach.
The pQCD predictions for the branching ratios of considered decays are:
$BR(B_d^0 \to J/\Psi \eta) = ( 1.96 ^{+9.68}_{-0.65} ) \times 10^{-6}$, which is consistent with
the first experimental measurement within errors;
while $BR(B_d^0 \to J/\Psi \eta') = (  1.09 ^{+3.76}_{-0.25}) \times 10^{-6}$, very
similar with $B_d^0 \to \jpsi \eta$ decay and can be tested by the forthcoming LHC experiments.
The measurements of these decay channels may help us to understand the QCD dynamics in
the corresponding energy scale, especially the reliability of pQCD approach to these kinds of
B meson decays.
\end{abstract}

\pacs{13.25.Hw, 12.38.Bx, 14.40.Nd}

\maketitle

Very recently,  the first observation of $B_d^0 \to  J/\Psi \eta$ decay was reported by Belle Collaboration
\cite{prl98}, and the branching ratio measured is
\beq
BR(B_d^0 \to J/\Psi \eta) &=& (9.5 \pm 1.7 (stat) \pm 0.8 (syst)) \times 10^{-6},
\label{eq:exp1}
\eeq
which is consistent with the currently available theoretical predictions \cite{prl98,plb318,jhep01}.

Up to now, the theoretical calculations for the branching ratios of $B_d \to J/\Psi \etap$ decays
were obtained by using the heavy quark factorization approximation in Ref.~\cite{plb318}, or
from the measured $J/\Psi \pi^0$ and $J/\Psi K^0$ branching ratios\cite{jhep01,pdg2006,hfag}
based on the assumption of the $SU(3)$ flavor symmetry of strong interaction.
In this paper, we will calculate the branching ratios of $B_d^0 \to J/\Psi \eta$ and
$B_d^0 \to J/\Psi \eta^{(\prime)}$ decays directly by employing the low energy effective Hamiltonian
\cite{buras96} and the
perturbative QCD (pQCD) factorization approach \cite{lb80,cl97,li2003}.

The paper is organized as follows: we present the formalism used in the
calculation of $B_d^0 \to J/\psi \eta^{(\prime)}$ decays in Sec.~\ref{sec:1}.
In Sec.~\ref{3}, we show the numerical results and compare them with the measured values.
A short summery and some conclusions are also included in this section.

\section{Formalism and Perturbative Calculations}\label{sec:1}

The pQCD approach has been developed earlier from the QCD hard-scattering approach \cite{lb80}, and has
been used frequently to calculate various B meson decay channels \cite{lb80,cl97,li2003,xiao06}.
For two body charmless hadronic $B_{d,s} \to M  \etap$ (here $M$ stands
for the pseudo-scalar or vector light mesons composed of the light quarks $u,d,s$) decays,
the pQCD predictions generally agree well with the measured values \cite{li2003,xiao06,ali07}.

In Refs.~\cite{ll03,llx05}, the authors calculated $B\to D_s^* K, D_s^{(*)+} D_s^{(*)-}$ and
$B_s \to  D^{(*)+} D^{(*)-}$ decays and found that the pQCD approach works well for such decays.
Here we try to apply the pQCD approach to calculate the B meson decays
involving the heavier $J/\Psi$ meson as one of the two final state mesons.

\subsection{Formulism}

In  pQCD approach, the decay amplitude of $B \to J/\Psi  P$ ($P=\eta, \eta^{(\prime)}$ here) decay
can bo written conceptually as the convolution,
\beq
{\cal A}(B \to M_1 M_2)\sim \int\!\! d^4k_1 d^4k_2 d^4k_3\ \mathrm{Tr} \left [ C(t) \Phi_{B}(k_1)
\Phi_{\jpsi}(k_2) \Phi_{P}(k_3) H(k_1,k_2,k_3, t) \right ],
\label{eq:con1}
\eeq
where the term ``$\mathrm{Tr}$" denotes
the trace over Dirac and color indices. $C(t)$ is the Wilson coefficient which
results from the radiative corrections at short distance. In the
above convolution, $C(t)$ includes the harder dynamics at larger
scale than $M_{B}$ scale and describes the evolution of local
$4$-Fermi operators from $m_W$ (the $W$ boson mass) down to
$t\sim\mathcal{O}(\sqrt{\bar{\Lambda} M_{B}})$ scale, where
$\bar{\Lambda}\equiv M_{B} -m_b$. The function $H(k_1,k_2,k_3,t)$
is the hard part and can be  calculated perturbatively. The
function $\Phi_M$ is the wave function which describes
hadronization of the quark and anti-quark to the meson $M$. While
the function $H$ depends on the process considered, the wave
function $\Phi_M$ is independent of the specific process. Using
the wave functions determined from other well measured processes,
one can make quantitative predictions here.

Using the light-cone coordinates the $B$ meson and the two final
state meson momenta can be written as
\beq
P_1 =\frac{M_{B}}{\sqrt{2}} (1,1,{\bf 0}_T), \quad
     P_2 =\frac{M_{B}}{\sqrt{2}} (1,r^2,{\bf 0}_T), \quad
     P_3 =\frac{M_{B}}{\sqrt{2}} (0,1-r^2,{\bf 0}_T),
\eeq
respectively, where $r=M_{J/\Psi}/M_{B}$, and the light meson
masses $m_\etap$ have been neglected.
The longitudinal polarization vector of the $J/\Psi$ meson, $\epsilon_L$,
is given by $\epsilon_L = \frac{M_B}{\sqrt{2}M_{J/\Psi}} (1, -r^2,{\bf 0}_T)$.
Putting the light (anti-) quark momenta in $B$, $J/\Psi$ and $\eta^{(')}$
mesons as $k_1$, $k_2$, and $k_3$, respectively, we can choose
\beq
k_1 = (x_1 P_1^+,0,{\bf k}_{1T}), \quad
k_2 = (x_2 P_2^+,0,{\bf k}_{2T}), \quad
k_3 = (0, x_3 P_3^-,{\bf k}_{3T}).
\eeq
Then, for $B \to J/\Psi \eta$ decay for example, the integration over
$k_1^-$, $k_2^-$, and $k_3^+$ in eq.(\ref{eq:con1}) will lead to
\beq
{\cal A}(B \to J/\Psi \eta^\prime) &\sim &\int\!\! d x_1 d x_2 d x_3 b_1 d b_1 b_2 d b_2
b_3 d b_3 \non && \cdot \mathrm{Tr} \left [ C(t) \Phi_{B}(x_1,b_1)
\Phi_{J/\Psi}(x_2,b_2) \Phi_{\eta}(x_3, b_3) H(x_i, b_i, t)
S_t(x_i)\, e^{-S(t)} \right ], \label{eq:a2}
\eeq
where $b_i$ is the conjugate space coordinate of $k_{iT}$, and $t$ is the largest
energy scale in function $H(x_i,b_i,t)$. The large logarithms $\ln
(m_W/t)$ are included in the Wilson coefficients $C(t)$. The large
double logarithms ($\ln^2 x_i$) on the longitudinal direction are
summed by the threshold resummation ~\cite{li02}, and they lead to
$S_t(x_i)$ which smears the end-point singularities on $x_i$. The
last term, $e^{-S(t)}$, is the Sudakov form factor which
suppresses the soft dynamics effectively ~\cite{soft}.
Thus it makes the perturbative calculation of the hard part $H$ applicable
at intermediate scale, i.e., $M_B$ scale. We will calculate
analytically the function $H(x_i,b_i,t)$ for the considered decays
in the first order in $\alpha_s$ expansion and give the convoluted
amplitudes in next section.

\begin{figure}[t,b]
\vspace{-2 cm} \centerline{\epsfxsize=21 cm \epsffile{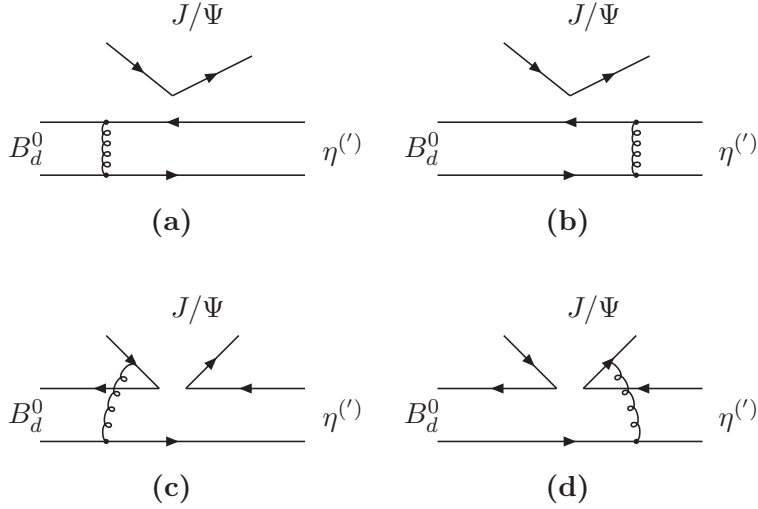}}
\vspace{-21cm} \caption{ Typical Feynman diagrams contributing to
the Cabibbo- and color- suppressed $B_{d}^0 \to J/\Psi\eta^{(')}$
decays.}
 \label{fig:fig1}
\end{figure}

\subsection{ The $B_d^0 \to J/\Psi \eta^{(')}$ Decays}\label{ssec:w-c}

The low energy effective Hamiltonian for decay modes $B_d^0 \to J/\psi
\eta^{(')}$ can be written as
\beq
\label{heff}
{\cal H}_{eff} = \frac{G_{F}} {\sqrt{2}} \, \left[ V_{cb} V_{cd}^* \left
(C_1(\mu) O_1^c(\mu) + C_2(\mu) O_2^c(\mu) \right) \right] \; ,
\eeq
with the four-fermion operators
\beq
\begin{array}{llllll}
O_1^{c} & = &  \bar d_\alpha\gamma^\mu (1 - \gamma_5) c_\beta\cdot
\bar c_\beta\gamma_\mu (1 - \gamma_5) b_\alpha\ , &O_2^{c} & =
&\bar d_\alpha\gamma^\mu (1 - \gamma_5) c_\alpha\cdot \bar
c_\beta\gamma_\mu (1 - \gamma_5) b_\beta\,  \label{eq:operators}
\end{array}
\eeq
where the Wilson coefficients $C_i(\mu)$ ($i=1,2$), we will
use the leading order (LO) expressions, although the
next-to-leading order (NLO) results already exist in the literature ~\cite{buras96}.
This is the consistent way to cancel the explicit $\mu$ dependence in the theoretical formulae. For the
renormalization group evolution of the Wilson coefficients from
higher scale to lower scale, we use the formulae as given in Ref.\cite{luy01} directly.

As for $B$ meson wavefunction, we make use of the same
parameterizations as used in the studies of different processes
\cite{luy01}. For vector $J/\psi$ meson, in terms of
the notation in Ref.~\cite{TLS}, we decompose the nonlocal matrix
elements for the longitudinally and transversely polarized $J/\psi$ mesons into
\beq
\Phi_{J/\Psi}(x) &=&\frac{1}{\sqrt{2N_c}} \bigg\{m_{J/\psi}\epsl_L
\Psi^L(x)+\epsl_L \psl \Psi^{t}(x) \bigg\}\;,
\eeq
Here, $\Psi^L$ denote for the twist-2 distribution amplitudes, and $\Psi^t$ for
the twist-3 distribution amplitudes. $x$ represents the momentum
fraction of the charm quark inside the charmonium.

The $J/\psi$ meson asymptotic distribution amplitudes read as
\cite{BC04}
\beq
\Psi^L(x)&=&9.58\frac{f_{J/\psi}}{2\sqrt{2N_c}}x(1-x)
\left[\frac{x(1-x)}{1-2.8x(1-x)}\right]^{0.7}\;,\nonumber\\
\Psi^t(x)&=&10.94\frac{f_{J/\psi}}{2\sqrt{2N_c}}(1-2x)^2
\left[\frac{x(1-x)}{1-2.8x(1-x)}\right]^{0.7}\;.\label{jda}
\eeq
It is easy to see that both the twist-2 and twist-3 DAs vanish at the end points
due to the factor $[x(1-x)]^{0.7}$.

From the effective Hamiltonian (\ref{heff}), the Feynman diagrams
corresponding to the considered decay are shown in Fig.1. With
the meson wave functions and Sudakov factors, the hard amplitude
is given as
\beq
F_{e\eta}&=& 8\pi C_F m_{B}^4\int_0^1 d x_{1} dx_{3}\, \int_{0}^{\infty} b_1 db_1 b_3 db_3\, \phi_{B}(x_1,b_1)
\non & & \times \left\{ ( 1-r^2 ) \left[(1+x_3 ( 1-r^2 ))
\phi_\eta^A(x_3, b_3) + r_0 (1-2x_3) \right.\right. \non &&
\left.\left. \cdot \phi_\eta^P(x_3,b_3)\right]+ r_0 \left[( 1 - 2
x_3 )+ r^2 (1 + 2 x_3 ) \right]\phi_\eta^T(x_3,b_3) \right. \non
&& \left. \cdot \alpha_s(t_e^1)\,
h_e(x_1,x_3,b_1,b_3)\exp[-S_{ab}(t_e^1)] \right.\non && \left. + 2
r_0 \left[1-(1-x_1) r^2 \phi_\eta^P (x_3, b_3)-x_1 r^2
\phi_\eta^A(x_3, b_3) \right] \right. \non && \left. \cdot
\alpha_s(t_e^2)h_e(x_3,x_1,b_3,b_1)\exp[-S_{ab}(t_e^2)] \right\}.
\label{eq:ab}
\eeq
where $r_0=m_0^\eta/m_B$; $C_F=4/3$ is a color factor. The function $h_e$, the scales
$t_e^i$ and the Sudakov factors $S_{ab}$ are displayed in Appendix \ref{sec:app1}.

For the non-factorizable diagrams 1(c) and 1(d), all three meson
wave functions are involved. The integration of $b_3$ can be
performed using $\delta$ function $\delta(b_3-b_1)$, leaving only
integration of $b_1$ and $b_2$. For the concerned operators, the corresponding decay amplitude
is
\beq
 M_{e\eta}&=& \frac{16 \sqrt{6}}{3}\pi C_F m_{B}^4
\int_{0}^{1}d x_{1}d x_{2}\,d x_{3}\,\int_{0}^{\infty} b_1d b_1
b_2d b_2\, \pb(x_1,b_1) \non
 & &\times
\left \{2 r r_c \phi_{J/\Psi}^t(x_2,b_2)\phi_\eta^A(x_3,b_2)-4 r
r_0 r_c \phi_{J/\Psi}^t(x_2,b_2)\phi_\eta^T(x_3,b_2)\right. \non
&& \left.-\left[x_2 r^2 + x_3 (1-2
r^2)\right]\phi_{J/\Psi}^L(x_2,b_2) \phi_\eta^A(x_3,b_2)
\right.\non
 & & \left. +2 \left[x_3 r_0+ (x_2-x_3)r_0 r^2\right]\phi_{J/\Psi}^L(x_2,b_2)\phi_\eta^T(x_3,b_2)
 \right. \non &&\cdot \alpha_s(t_f) h_f(x_1,x_2,x_3,b_1,b_2)\exp[-S_{cd}(t_f)] \} \; \label{eq:cd}.
\eeq where $r_c=m_{c}/m_{B}$,$m_c$ is the mass for $c$ quark.

For the $B_d^0 \to J/\Psi \eta^{\prime}$ decay, the Feynman
diagrams are obtained by replacing the $\eta$ meson in Fig.~1 with the meson $\etapr$.
The corresponding expressions of decay amplitudes will be similar with those as given in
Eqs.(\ref{eq:ab}-\ref{eq:cd}), since the $\eta$ and $\etapr$ are
all light pseudoscalar mesons and have the similar wave functions.
The expressions of $B_d^0 \to J/\Psi \eta^{\prime}$ decay can be
obtained simply by the following replacements
\beq
\pe \longrightarrow \phi_{\etapr}^A, \quad \pep\longrightarrow \phi^P_{\etapr}, \quad
\pet \longrightarrow \phi^T_{\etapr}, \quad r_0 \longrightarrow r'_0.
\eeq


For the $\eta-\eta^\prime$ system, there exist two popular mixing basis:
the octet-singlet basis and the quark-flavor basis \cite{fk98,0501072}.
Here we  use  the quark-flavor basis \cite{fk98} and define
\beq
\eta_q=(u\bar{u} + d\bar{d})/\sqrt{2},
\qquad \eta_s=s\bar{s}. \label{eq:qfbasis}
\eeq
The physical states $\eta$ and $\eta^\prime$ are related to $\eta_q$ and
$\eta_s$ through a single mixing angle $\phi$,
\beq
\left(
\begin{array}{c} \eta\\ \eta^\prime \\ \end{array} \right ) &=&
U(\phi) \left( \begin{array}{c}
 \eta_q\\ \eta_s \\ \end{array} \right ) =
  \left( \begin{array}{cc}
 \cos{\phi} & -\sin{\phi} \\
 \sin{\phi} & \cos\phi \end{array} \right )
 \left( \begin{array}{c}  \eta_q\\ \eta_s \\ \end{array} \right ).
 \label{eq:e-ep}
\eeq
The three input parameters $f_q$, $f_s$ and $\phi$ in the
quark-flavor basis have been extracted from various related
experiments \cite{fk98,0501072}
\beq
f_q = (1.07\pm 0.02) f_\pi, \quad f_s = (1.34 \pm 0.06) f_\pi, \quad \phi = 39.3^\circ \pm
1.0^\circ, \label{eq:e}
\eeq
where $f_\pi=130$ MeV. In the numerical calculations, we will use these mixing parameters as
inputs. It worth of mentioning that the effects of possible gluonic component of $\eta^\prime$
meson will not considered here since it is small in size \cite{xiao06,0609165,0703187}.


For $B_d^0 \to J/\Psi  \eta$ decay, by combining the contributions
from different diagrams, the total decay amplitude can be written
as
\beq
{\cal M}(B_d^0 \to J/\Psi \eta) &=& V_{cb}V_{cd}^*
F_1(\phi) \left\{ F_{e\eta} f_{J/\Psi}
 \left[ \left( C_1 +
\frac{1}{3}C_2\right)\right ]
 + M_{e\eta} C_2 \right\}  \label{eq:m1}
\eeq
where the relevant mixing parameter is $F_1(\phi) = \cos{\phi}/\sqrt{2}$.

It should be mentioned that the Wilson coefficients $C_i=C_i(t)$ in Eq.~(\ref{eq:m1}) should be
calculated at the appropriate scale $t$ using equations as given
in the Appendices of Ref.~\cite{luy01}. Here the scale $t$ in the
Wilson coefficients should be taken as the same scale appeared in
the expressions of decay amplitudes in Eqs.~(\ref{eq:ab}) and
(\ref{eq:cd}). This is the way in pQCD approach to eliminate the
scale dependence. In order to estimate the effect of higher order
corrections, however, we introduce a scale factor $a_t=1.0 \pm 0.2$ and vary the scale $t_{max}$ as described
in Appendix A.

Similarly, the decay amplitudes for $B_d^0 \to J/\Psi
\eta^{\prime}$ decay can be obtained easily from Eq.(\ref{eq:m1})
by the following replacements of $F_1(\phi) \to F^{\prime}_1(\phi)=\sin{\phi}/\sqrt{2}$.

\section{Numerical results and Discussions}\label{3}

In this section, we will calculate the branching ratios for those
considered decay modes. The input parameters and the wave
functions to be used are given in Appendix \ref{sec:app2}. In
numerical calculations, central values of input parameters will be
used implicitly unless otherwise stated.

With the complete decay amplitudes, we can obtain the decay width
for the considered decays,
\beq
 \Gamma(B_d^0 \to J/\psi \eta^{(')}) = \frac{G_F^2 M_B^3}{32\pi}(1-r^2)
\left| {\cal M} (B_d^0 \to J/\psi \eta^{(')})\right|^2.
\eeq

By employing the quark-flavor scheme of $\eta-\eta^\prime$ system
and using the mixing parameters as given in Eq.~(\ref{eq:e}), one
finds the branching ratios for the considered two decays with
error bars as follows:
 \beq Br(\ B_d^0 \to J/\Psi
\eta) &=& \left [1.96^{+0.71}_{-0.50}(\omega_b)
 ^{+9.65}_{-0.39}(a_t)^{+0.32}_{+0.13}(a_2)^{+0.14}_{-0.13}( f_{J/\Psi})\right ] \times 10^{-6}, \label{eq:brje1}\\
Br(\ B_d^0 \to J/\Psi \eta^{\prime}) &=& \left
[1.09^{+0.32}_{-0.24}(\omega_b)
 ^{+3.73}_{+0.01}(a_t)^{+0.28}_{+0.01}(a_2)^{+0.08}_{-0.07}( f_{J/\Psi})\right ] \times 10^{-6}, \label{eq:brjp1}
 \eeq
where the main errors are induced by the uncertainties of $\omega_b=0.40 \pm 0.05$ GeV,
$a_t = 1.0 \pm 0.2$, $a_2 = 0.115 \pm 0.115$ and $f_{J/\Psi}= 0.405 \pm 0.014$ GeV  , respectively.
One can see that the pQCD predictions are sensitive to the variations of $\omega_b$ and $a_t$.

For $B_d^0 \to J/\Psi \eta $ decay, the central value of the pQCD prediction for
$Br(B_d^0 \to J/\Psi \eta)$ is a factor of 4 smaller than the measured value as
given in Eq.~(\ref{eq:exp1}) \cite{prl98}.
But the pQCD prediction is in fact still consistent with
Belle's first measurement if we take the large theoretical and experimental errors into account.
By varying the scale factor $a_t$ in the range of $a_t=[0.8,1.0]$, for example, the central
value of $Br(B \to \jpsi \eta)$ will change in the range of $[0.2,1.1]\times 10^{-5}$ accordingly.
It is not difficult to understand such $a_t$ dependence. Since the $J/\Psi$ meson is much heavier than
light mesons, and therefore moving not as fast as those light meson when B meson is decaying.
So a small decrease of the scale $t_i$ will lead to a larger Wilson coefficients $C_{1,2}(t)$ and
$\alpha_s(t_i)$, and consequently results in a larger decay rate.

For $B_d^0 \to J/\Psi \eta^\prime$ decay, only experimental upper limit (at $90\%$ C.L) is available
now: $BR(B^0 \to  J/\Psi \eta^\prime )  <  6.3 \times 10^{-5}$ \cite{pdg2006,hfag}.
The pQCD prediction for the branching ratio of $B_d^0 \to J/\Psi \eta^\prime$ decay is very
similar in magnitude with that of $B_d^0 \to J/\Psi \eta$, consistent with the upper limit and
will be tested in the forthcoming LHC experiments.

At the leading order, only the tree Feynman diagrams as shown in Fig.~1 contribute to
$B_d^0 \to \jpsi \etap$ decays.
There exists no CP violation in these decays within the standard model, since there is
only one kind of Cabibbo-Kabayashi-Muskawa (CKM) phase involved in the corresponding decay amplitudes,
as can be seen from eq.~(\ref{eq:m1}).

In short,  we calculated the branching ratios of  $B_d^0 \to
J/\Psi \eta$ and $ B_d^0 \to J/\Psi \eta^{\prime}$ decays at the
leading order by using the pQCD factorization approach. Besides
the usual factorizable diagrams, the non-factorizable spectator diagrams are
also calculated analytically in the pQCD approach.
By keeping the transverse momentum $k_T$, the end-point singularity disappears in
our calculation.

From our calculations and phenomenological analysis, we found the
following results:
\begin{itemize}
\item
Using the quark-flavor scheme, the pQCD predictions for the
branching ratios are \beq Br(B_d^0 \to J/\Psi \eta)
&=& \left (1.96 ^{+9.68}_{-0.65} \right ) \times 10^{-6},\\
Br(B_d^0 \to J/\Psi \eta^\prime)&=& \left ( 1.09 ^{+3.76}_{-0.25}
\right ) \times 10^{-6},
\eeq
where the various errors as specified previously have been added in quadrature.

\item
The major theoretical errors of the pQCD predictions are
induced by the uncertainties of the hard energy scale $t_i$'s and the
parameters $\omega_b$.

\end{itemize}

\begin{acknowledgments}

X.~Liu would like to acknowledge the financial support of The Scientific Research
Start-up Fund of Zhejiang Ocean University under Grant No.21065010706.
This work was partially supported by the National Natural Science Foundation of China under Grant
No.10575052, and by the Specialized Research Fund for
the Doctoral Program of Higher Education (SRFDP) under Grant No.~20050319008.

\end{acknowledgments}


\begin{appendix}

\section{Related Functions }\label{sec:app1}

We show here the function $h_i$'s, coming from the Fourier
transformations  of the function $H^{(0)}$, \beq
 h_e(x_1,x_3,b_1,b_3)&=&
 K_{0}\left(\sqrt{x_1 x_3(1-r^2)} m_{B} b_1\right)
 \left[\theta(b_1-b_3)K_0\left(\sqrt{x_3(1-r^2)} m_{B}
b_1\right)\right.
 \non
& &\;\left. \cdot I_0\left(\sqrt{x_3(1-r^2)} m_{B}
b_3\right)+\theta(b_3-b_1)K_0\left(\sqrt{x_3(1-r^2)}  m_{B}
b_3\right)\right.
 \non
& &\;\left. \cdot I_0\left(\sqrt{x_3(1-r^2)}  m_{B}
b_1\right)\right] S_t(x_3), \label{he1} \eeq 
 \beq
 h_{f}(x_1,x_2,x_3,b_1,b_2) &=&
 \biggl\{\theta(b_2-b_1) \mathrm{I}_0(M_{B}\sqrt{x_1 x_3(1-r^2)} b_1)
 \mathrm{K}_0(M_{B}\sqrt{x_1 x_3(1-r^2)} b_2)
 \non
&+ & (b_1 \leftrightarrow b_2) \biggr\}  \cdot\left(
\begin{matrix}
 \mathrm{K}_0(M_{B} F_{(1)} b_2), & \text{for}\quad F^2_{(1)}>0 \\
 \frac{\pi i}{2} \mathrm{H}_0^{(1)}(M_{B}\sqrt{|F^2_{(1)}|}\ b_2), &
 \text{for}\quad F^2_{(1)}<0
\end{matrix}\right),
\label{eq:pp1}
 \eeq
where $J_0$ is the Bessel function, $K_0$ and $I_0$ are
the modified Bessel functions with  $K_0 (-i x) = -(\pi/2) Y_0 (x) + i
(\pi/2) J_0 (x)$, and $F_{(j)}$'s are defined by
\beq
F^2_{(1)}&=&(x_1 -x_2) x_3(1-r^2)+r_c^2\;,\\
F^2_{(2)}&=&(x_1 -x_2) x_3(1-r^2)+r_c^2\; .
 \eeq
The threshold resummation form factor $S_t(x_i)$ is adopted from
Ref.~\cite{TLS}
\beq
S_t(x)=\frac{2^{1+2c} \Gamma (3/2+c)}{\sqrt{\pi} \Gamma(1+c)}[x(1-x)]^c,
\eeq
where the parameter $c=0.3$. This function is normalized to unity.

The Sudakov factors used in the text are defined as
\beq
S_{ab}(t) &=& s\left(x_1 m_{B}/\sqrt{2}, b_1\right) +s\left(x_3
m_{B}/\sqrt{2}, b_3\right) +s\left((1-x_3) m_{B}/\sqrt{2},
b_3\right) \non
&&-\frac{1}{\beta_1}\left[\ln\frac{\ln(t/\Lambda)}{-\ln(b_1\Lambda)}
+\ln\frac{\ln(t/\Lambda)}{-\ln(b_3\Lambda)}\right],
\label{wp}\\
S_{cd}(t) &=& s\left(x_1 m_{B}/\sqrt{2}, b_1\right)
 +s\left(x_2 m_{B}/\sqrt{2}, b_2\right)
+s\left((1-x_2) m_{B}/\sqrt{2}, b_2\right) \non
 && +s\left(x_3
m_{B}/\sqrt{2}, b_1\right) +s\left((1-x_3) m_{B}/\sqrt{2},
b_1\right) \non
 & &-\frac{1}{\beta_1}\left[2
\ln\frac{\ln(t/\Lambda)}{-\ln(b_1\Lambda)}
+\ln\frac{\ln(t/\Lambda)}{-\ln(b_2\Lambda)}\right], \label{Sc}
\eeq
where the function $s(q,b)$ are defined in the Appendix A of
Ref.~\cite{luy01}. The scale $t_i$'s in the above equations are
chosen as
\beq
t_{e}^1 &=& a_t \cdot  {\rm max}(\sqrt{x_3(1-r^2)}M_{B},1/b_1,1/b_3), \non
t_{e}^2 &=& a_t \cdot  {\rm max}(\sqrt{x_1(1-r^2)}M_{B},1/b_1,1/b_3),\non
t_{f} &=& a_t \cdot  {\rm max}(\sqrt{x_1 x_3(1-r^2)}M_{B},\sqrt{(x_1-x_2)x_3(1-r^2)+r_c^2}M_{B},
1/b_1,1/b_2),
\eeq
where $a_t=1.0\pm 0.2$ and $r=M_{\jpsi}/M_B$. The scale $t_i$'s  are chosen as the maximum energy scale
appearing in each diagram to kill the large logarithmic radiative corrections.

\section{Input parameters and wave functions} \label{sec:app2}

The masses, decay constants, QCD scale  and $B_d^0$ meson lifetime
are
\beq
 \Lambda_{\overline{\mathrm{MS}}}^{(f=4)} &=& 250 {\rm MeV}, \quad
 f_\pi = 130 {\rm MeV}, \quad  f_{J/\Psi} = 405 {\rm MeV}, \non
 m_0^{\eta_{d\bar{d}}}&=& 1.08 {\rm GeV},\quad M_{B_d^0} = 5.28  {\rm MeV}, \quad
  M_{J/\Psi} = 3.097 {\rm GeV}, \non
 M_W &=& 80.41{\rm
 GeV}, \quad \tau_{B_d^{0}}=1.54\times10^{-12}{\rm s}.
 \label{para}
\eeq

For the CKM matrix elements, here we adopt the Wolfenstein
parametrization for the CKM matrix, and take $\lambda=0.2272,
A=0.818, \rho=0.221$ and $\eta=0.340$ \cite{pdg2006}.

For the $B$ meson wave function, we adopt the model \beq
\phi_{B}(x,b) &=& N_{B} x^2(1-x)^2 \mathrm{exp} \left
 [ -\frac{M_{B}^2\ x^2}{2 \omega_{b}^2} -\frac{1}{2} (\omega_{b} b)^2\right],
 \label{phib}
\eeq where $\omega_{b}$ is a free parameter and we take
$\omega_{b}=0.40\pm 0.05$ GeV in numerical calculations, and
$N_{B}=91.745$ is the normalization factor for $\omega_{b}=0.40$
for the $B$ meson.

The wave function for $d\bar{d}$ components of $\eta^{(\prime)}$ meson is  given by
\beq
\Phi_{\eta_{d\bar{d}}}(p,x,\zeta)\equiv \frac{i
\gamma_5}{\sqrt{2N_c}} \left [ \psl
\phi_{\eta_{d\bar{d}}}^{A}(x)+m_0^{\eta_{d\bar{d}}}
\phi_{\eta_{d\bar{d}}}^{P}(x)+\zeta m_0^{\eta_{d\bar{d}}} ( \vsl
\nsl - v\cdot n)\phi_{\eta_{d\bar{d}}}^{T}(x) \right ],
\label{eq:ddbar}
\eeq
where $p$ and $x$ are the momentum and the
momentum fraction of $\eta_{d\bar{d}}$ respectively, while
$\phi_{\eta_{d\bar{d}}}^A$, $\phi_{\eta_{d\bar{d}}}^P$ and
$\phi_{\eta_{d\bar{d}}}^T$ represent the axial vector,
pseudoscalar and tensor components of the wave function
respectively.  We here assume that the wave function of $\eta_{d\bar{d}}$ is same as the $\pi$
wave function based on SU(3) flavor symmetry.
The parameter $\zeta$ is either $+1$ or $-1$ depending on the assignment of the
momentum fraction $x$.

The explicit expression of chiral enhancement scale
$m_0^{q}=m_0^{\eta_{d\bar{d}}}$ is given by \cite{0609165}
\beq
m_0^q\equiv \frac {m_{qq}^2}{2m_q}&=&
\frac{1}{2m_q}[m_{\eta}^2\cos^2{\phi}+m_{\eta^\prime}^2\sin^2{\phi}
-\frac{\sqrt{2}f_s}{f_q}(m_{\eta^\prime}^2-m_{\eta}^2)\cos{\phi}sin{\phi}], \label{eq:m0q}
\eeq
and numerically $ m_0^q= 1.07 {\rm MeV}$ for $m_\eta=547.5$ MeV,
$m_{\eta^\prime}=957.8$ MeV, $f_q=1.07f_\pi$, $f_s=1.34 f_\pi$ and $\phi=39.3^\circ$.

For the distribution amplitude $\phi_{\eta_{q}}^A$,
$\phi_{\eta_{q}}^P$ and $\phi_{\eta_{q}}^T$, we utilize the
results for $\pi$ meson obtained from the light-cone sum
rule~\cite{ball} including twist-3 contributions:
\beq
\phi_{\eta_{q}}^A(x)&=&\frac{3}{\sqrt{2N_c}}f_{q}x(1-x)
\left\{ 1+a_2^{\eta_{q}}\frac{3}{2}\left [5(1-2x)^2-1 \right
]\right. \non &&\left.
+ a_4^{\eta_{q}}\frac{15}{8} \left [21(1-2x)^4-14(1-2x)^2+1 \right ]\right \},  \\
\phi^P_{\eta_{q}}(x)&=&\frac{1}{2\sqrt{2N_c}}f_{q} \left \{
1+ \frac{1}{2}\left (30\eta_3-\frac{5}{2}\rho^2_{\eta_{q}}
\right ) \left [ 3(1-2x)^2-1 \right] \right.  \non && \left. +
\frac{1}{8}\left (
-3\eta_3\omega_3-\frac{27}{20}\rho^2_{\eta_{q}}-
\frac{81}{10}\rho^2_{\eta_{q(s)}}a_2^{\eta_{q}} \right )
\left [ 35 (1-2x)^4-30(1-2x)^2+3 \right ] \right\}, \ \ \\
\phi^T_{\eta_{q}}(x) &=&\frac{3}{\sqrt{2N_c}}f_{q} (1-2x)
\non
 && \cdot \left [ \frac{1}{6}+(5\eta_3-\frac{1}{2}\eta_3\omega_3-
\frac{7}{20}\rho_{\eta_{q}}^2
-\frac{3}{5}\rho^2_{\eta_{q}}a_2^{\eta_{q,s}})(10x^2-10x+1)\right ],
\eeq
with the updated Gegenbauer moments \cite{ball06}
\beq
a^{\eta_{q}}_2&=&0.115, \quad  a^{\eta_{q}}_4 =-0.015,
\quad  \rho_{\eta_q}=2m_{q}/m_{qq}, \quad \eta_3=0.015, \quad
\omega_3=-3.0.
 \eeq

\end{appendix}


\end{document}